\documentclass{PoS}

\title{Neutrino Telescope in Lake Baikal: Present and Future}

\ShortTitle{Baikal GVD Neutrino Telescope: Present and Future}

\author{A.D.~Avrorin$^{a}$, A.V.~Avrorin$^a$, V.M.~Aynutdinov$^a$, R.~Bannash$^g$, I.A~Belolaptikov$^b$, V.B.~Brudanin$^b$, N.M.~Budnev$^c$, G.V.~Domogatsky$^a$, A.A.~Doroshenko$^a$, R.~Dvornick\'y$^{b,h}$, A.N.~Dyachok$^c$, Zh.-A.M.~Dzhilkibaev$^a$, L. Fajt$^{h,i}$, S.V~Fialkovsky$^e$, A.R.~Gafarov$^c$, K.V.~Golubkov$^a$, N.S.~Gorshkov$^b$, T.I.~Gress$^c$, R.~Ivanov$^b$, K.G.~Kebkal$^g$, O.G.~Kebkal$^g$, E.V.~Khramov$^b$ , M.M.~Kolbin$^b$, K.V.~Konischev$^b$, A.V.~Korobchenko$^b$, A.P.~Koshechkin$^a$, A.V.~Kozhin$^d$, M.V.~ Kruglov$^b$, M.K.~Kryukov$^a$, V.F.~Kulepov$^e$, M.B.~Milenin$^a$, R.A.~Mirgazov$^c$, V.~Nazari$^b$, \fbox{A.I.~Panfilov$^a$}, D.P.~Petukhov$^a$ E.N.~Pliskovsky$^b$, M.I.~Rozanov$^f$, E.V.~Rjabov$^c$, V.D.~ Rushay$^b$, G.B.~Safronov$^b$, B.A.~Shaybonov$^b$, M.D.~Shelepov$^a$, \speaker{F.~\v{S}imkovic}$^{,b,h,i}$, A.V.~Skurikhin$^d$, A.G.~Solovjev$^b$, M.N.~ Sorokovikov$^b$, I.~\v{S}tekl$^i$, O.V.~Suvorova$^a$, E.O.~Sushenok$^b$, V.A.~Tabolenko$^c$, B.A.~Tarashansky$^c$, S.A.~Yakovlev$^g$\\
$^a$ Institute for Nuclear Research, Russian Academy of Sciences, Moscow, 117312 Russia\\
$^b$ Joint Institute for Nuclear Research, Dubna, 141980 Russia\\
$^c$ Irkutsk State University, Irkutsk, 664003 Russia\\
$^d$ Institute of Nuclear Physics, Moscow State University, Moscow, 119991 Russia\\
$^e$ Nizhni Novgorod State Technical University, Nizhni Novgorod, 603950 Russia\\
$^f$ St. Petersburg State Marine Technical University, St. Petersburg, 190008 Russia\\
$^g$ EvoLogics Gmbh, Germany\\ 
$^h$ Comenius University, Mlynska Dolina F1, Bratislava, 842 48 Slovakia\\
$^i$ Czech Technical University in Prague, Prague, 128 00 Czech Republic\\
E-mail: \email{simkovic@fmph.uniba.sk}
}

\abstract{A significant progress in the construction and operation of the Baikal Gigaton Volume Detector
in Lake Baikal, the largest and deepest freshwater lake in the world, is reported.
The effective volume of the detector for neutrino initiated cascades of relativistic particles
with energy above 100 TeV has been increased up to about 0.25 km$^3$. This unique scientific facility,
the largest operating neutrino telescope in Northern Hemisphere, 
allows already to register two to three events per year from astrophysical neutrinos with energies
exceeding 100 TeV. Preliminary results obtained with data recorded in 2016-2018 are announced.
Multimessenger approach is used to relate finding of cosmic neutrinos  with those of classical astronomers,
with X-ray or gamma-ray observations and the gravitational wave events. 
}

\FullConference{36th International Cosmic Ray Conference -ICRC2019-\\
		July 24th - August 1st, 2019\\
		Madison, WI, U.S.A.}

\begin{document}

\section{Introduction}

High-energy neutrino astronomy, a very recent and lively research field, has emerged by
a construction of gigaton volume neutrino telescopes deep in ice and under water at both
the Southern (IceCube) and Northern (Baikal-GVD, KM3NeT) Hemispheres.
Its primary goal is to study the most distant, powerful energy sources in the universe
using high-energy neutrinos.
The subject of interest are astrophysical sources, namely the collision and merger of
a pair of neutron stars, or a neutron star with a black hole, active galactic nuclei
powered by mass accretion onto black hole at the center of its host galaxy, etc.
They can be observed by a detection of astrophysical neutrinos in the TeV-PeV range coming
from outside our galaxy. The high energy universe is a great laboratory to search also for
fundamental properties of matter, in particular dark matter particles, axions, magnetic
monopoles, new light bosons etc.

Recently, IceCube telescope confirmed that blazars, an intense
source of radiation powered by a supermassive black hole, produce ultra-high energy neutrinos \cite{ricap2}.
A 290 TeV neutrino was registered that could be traced back to TXS 0506+056 blazar that
emits copious amounts of gamma rays. In this way the era of multimessenger astronomy
has started allowing to look at sky simultaneously with light, neutrinos, and gravitational waves.

In this contribution a significant progress in construction and performance of underwater neutrino
telescope in Lake Baikal is reviewed. A description of the Baikal-GVD telescope working
principles as well as the reconstruction of high-energy events which are identified above
the background of atmospheric neutrinos are shortly presented \cite{gvdepj19}.

\section{The Baikal-GVD neutrino telescope}

Next generation cubic kilometer scale neutrino telescope Baikal-GVD is currently
under construction in Lake Baikal since 2015. The detector is specially designed
for search for high energies neutrinos whose sources are not yet reliably identified.

The  Baikal-GVD detector is  3  dimensional  lattice  of optical modules (OMs)
that house photomultiplier tubes to detect Cherenkov light from secondary muons
and cascades produced in neutrino interactions in the water of the lake.
The OMs are arranged on vertical load-carrying  cables  to  form  strings.  The
configuration  of  telescope  consists  of  functionally independent clusters  of
strings,  which  are  connected  to  shore  by  individual  electro-optical cables.
Each cluster comprises in  eight strings 288 optical modules arranged at depths
down to 1300 m. Seven  peripheral  strings  are  uniformly  located  at
a  60  m  distance around  a  central  one. OMs on each string  are  combined  in  sections,
detection  units  of  telescope.  The  distances  between  the central strings
of neighboring clusters are about 300 m.

A basic element of operation of the Baikal-GVD detector is a section of 12 OMs
distributed vertically along the string, spaced by 15 m, and a central electronics module (CeM).
There are 3 sections of 36 OMs per string and 8 strings in cluster, each is an independently
working sub-array of the Baikal-GVD. As light sensor of OM a single photomultiplier tube (PMT) 
Hamamatsu R7081-100 with a 10-inch hemispherical photocathode and quantum efficiency
up to 35\% is exploited \cite{ricap8}. The PMT signals are amplified and transmitted
to the ADC units to measure the pulse shape with a sampling rate 200 MHz \cite{gvdiet14}.
Each ADC unit is equipped with quartz oscillators that allow measuring the arrival time of
photons. A global trigger signal is used to unify the event times measured by different ADC
units within the same cluster to a single time scale with accuracy $\sim$2 ns. The trigger condition is a coincidence
of pulses from any neighboring optical modules with thresholds $\sim$2 and $\sim$5 photoelectrons.
Once trigger condition is fulfilled, a request signal is generated to the cluster center module (CC)
through 1 km long line. CC generates an acknowledgement signal to all CMs in a cluster, when the request signal is
received. When CM is caught the signal, a timestamp is defined and the CM starts to form data. The
programmable gate array memory buffer allows to acquire the OM signal waveform for a time interval as long as 5 $\mu s$,
this information then is transferred to the shore station and stored as raw data in binary files. Such a
trigger system approach allows to have all signal waveforms on each channel in an event from only
one triggered pair of adjacent channels, while the data arriving from sections can be processed in
real time mode. All raw data, for an amount as large as 40 Gbyte per cluster per day, are transferred to the shore
station, stored there, and then transmitted at 5 Mbit/s on a 40km channel to Baikalsk (Irkutsk region),
where data are archived and transferred through a high-bandwidth internet to the central storage and
processing facility in the JINR in Dubna (Moscow region) every 4 hours.

\begin{figure}[!t]
\begin{center}$
\begin{array}{cc}
\includegraphics[width=60mm,height=60mm]{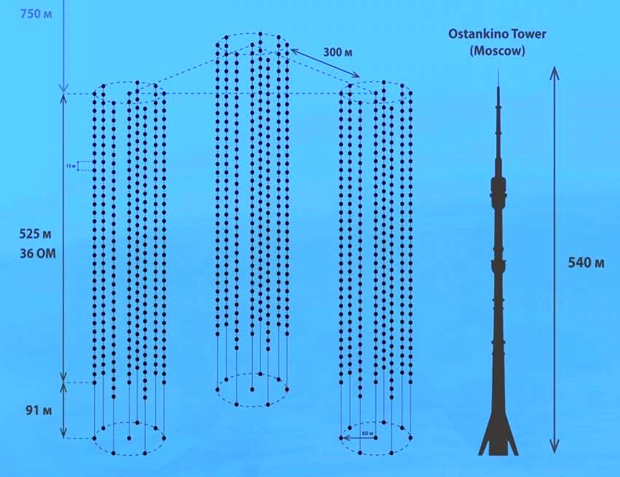} &
\includegraphics[width=80mm,height=60mm]{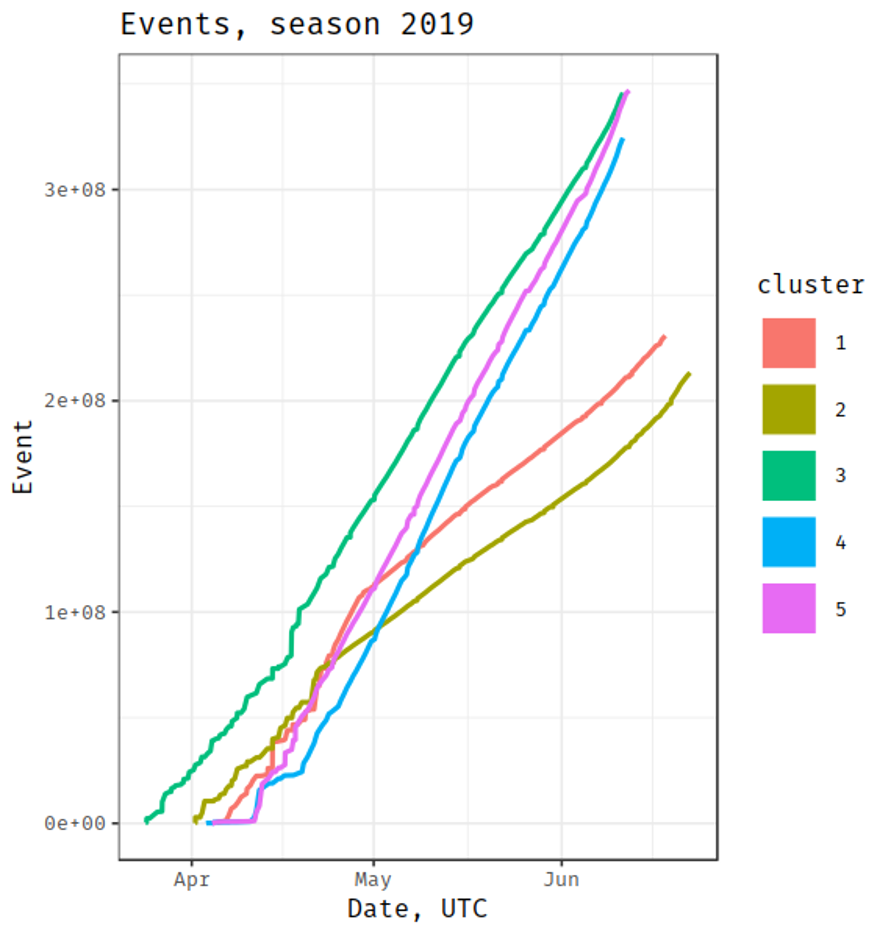} 
\end{array}$
\end{center}
\caption{Schematic drawing of the Baikal-GVD three clusters operated in 2018 (left panel) and
  integrated number of events recorded by the clusters from April to June 2019 (right panel).}
\label{fig.1}
\end{figure}

Pulse shape analysis allows determining the number of signals coming from the PMT, their charges
and registration times. On the basis of measurements of the time of registration of the pulses
and their charges on all triggered channels, the main parameters of an event are determined:
the direction and energy of muons and cascades, respectively. The accuracy of the
direction reconstruction of muon tracks amounts to 0.5$^o$, and that of the energy of cascades
to about 20\%. For amplitude and time calibration of the detector, LED calibration sources
are used. They are located both in the optical modules and in separate deep-water glass spheres (LED
matrix). Each OM comprises two LEDs, the light from which spreads up along the string. These LEDs
are employed for channel calibration within sections of OMs. The LED matrices are used for inter-section
time calibration. Each matrix comprises 12 synchronized LEDs flashers, pulses from which can be
detected by several sections. The accuracy of the time calibration is about 2 ns \cite{gvdshel18}.
Laser sources are used for inter-cluster calibration. The laser calibration source generates
light flashes with about $10^{15}$ photons at maximum power running. The laser wavelength is 532 nm,
the pulse duration about 1 ns. The laser is located approximately in the center of the group
of three clusters, at the level of the middle of the lower sections of the optical modules
(see Figure \ref{fig.1}, left panel). The laser is equipped with a diffuser that forms a quasi-isotropic light flux.
One laser source allows calibrating three clusters simultaneously. Details of data acquisition, basic controls,
methods of calibrations, hard-and soft-ware triggers can be found in \cite{gvdiet14} and \cite{ricap10}.

The first phase (Phase 1) of the construction of the Baikal-GVD telescope with 8 clusters
with 2304 light sensors in total covering 0.4 km$^3$ of instrumented volume is
expected to be finished in 2021. However, starting April 2015, when the first GVD cluster
Dubna was commissioned, data from the first clusters of the detector are already collected,
analyzed and exploited for a development of data analysis software.
Since April 2018 the telescope had been successfully
operated in complex of three functionally independent clusters i.e. sub-arrays
of optical modules (OMs) hosting 864 OMs on 24 vertical strings.
A significant amount of work has been accomplished during this year winter expedition
resulting in operation of two new clusters, altogether 5 clusters with
effective volume of 0.25 km$^3$, starting April 2019. In Fig. \ref{fig.1} (right panel) 
the event accumulation rate of the five clusters from April to June 2019 is presented. 
The change in slope for clusters 1 and 2 is due to raising the trigger threshold.
Their DAQ (already optimized for the last 3 clusters) was challenged by the high data rates.
About $1.5\times 10^9$ events have been recorded, with a data taking efficiency of ~90\%
for single clusters and almost 100\% with at least one cluster taking data.
The current effective volume of the Baikal-GVD detector is just 
one fourth of the size of the present world leader - the IceCube Neutrino Observatory
at the South Pole. A possible extension of Phase 1 to Phase 2 with construction
of additional 14 clusters will depend on the performance and physical output
of the Baikal GVD detector in 2021.

\section{On the search for high energy neutrino events and their origin}

To  priority tasks of the Baikal-GVD telescope belong a search for astrophysical neutrino sources
and neutrino events correlated in time and direction with sources of variable luminosity what
can be achieved by a study of tracks ascribed to  $\nu_\mu$, $\overline{\nu}_\mu$ and high-energy
cascade events ascribed to $\nu_e$, $\overline{\nu}_e$ and $\nu_{\tau}$, $\overline{\nu}_{\tau}$.
The reconstruction of Cherenkov radiation
being emitted by cascade particles and straight-line moving muons generated in the high energy
neutrino interactions with Baikal water allows to conclude about energies and direction
of incoming neutrinos. 
The Baikal-GVD telescope has modular structure what allows to collect and analyze data since 
April 2015, when the first GVD cluster was commissioned and the neutrino telescope has been in
a permanent operation.

\subsection{Detection of muon neutrinos}

A search for upward moving muon neutrinos having a collision with water environment
of the Baikal-GVD detector was performed with data collected by the first cluster of
the telescope in 2016. For analysis based on precise measurements of
optical module positions, timing and amplitude calibrations
a set of 70 runs in which detector demonstrated stable behavior was chosen.
The total exposition time of this sample of data is close to 33 days.

The collection of PMT signals in each event includes noise pulses due to PMT dark current
and to chemiluminescence background in the Baikal water.
Rates of noise impulses vary between 20 and 100 kHz depending on season and depth.
The noise pulses are rejected with help of the causality requirement:
$|t_i-t_j|\le \frac{\Delta R_{ij}}{c_w}+t_s$, where $t_i$, $t_j$ are impulse times,
$\Delta R_{ij}$ is distance between modules, $c_w$ is speed of light in water and $t_s = 10$ ns.
A simple track position and direction estimation is done for each causally-connected group.
OM hits which do not obey model of muon propagation and direct Cherenkov light emission are
excluded in a gradually tightening set of cuts on hit residuals. The set of hits selected
with this procedure has a noise contamination at the level of 1-2\% depending on the elevation
of the muon track. Selected set of pulses is used for the track fit under the assumption of
direct Cherenkov emission from the muon track. The resulting median resolution of the procedure
as measured in the up-going  neutrino Monte Carlo sample is at the level of 1 degree.

\begin{figure}[!t]
\begin{center}$
\begin{array}{cc}
\includegraphics[width=75mm,height=60mm]{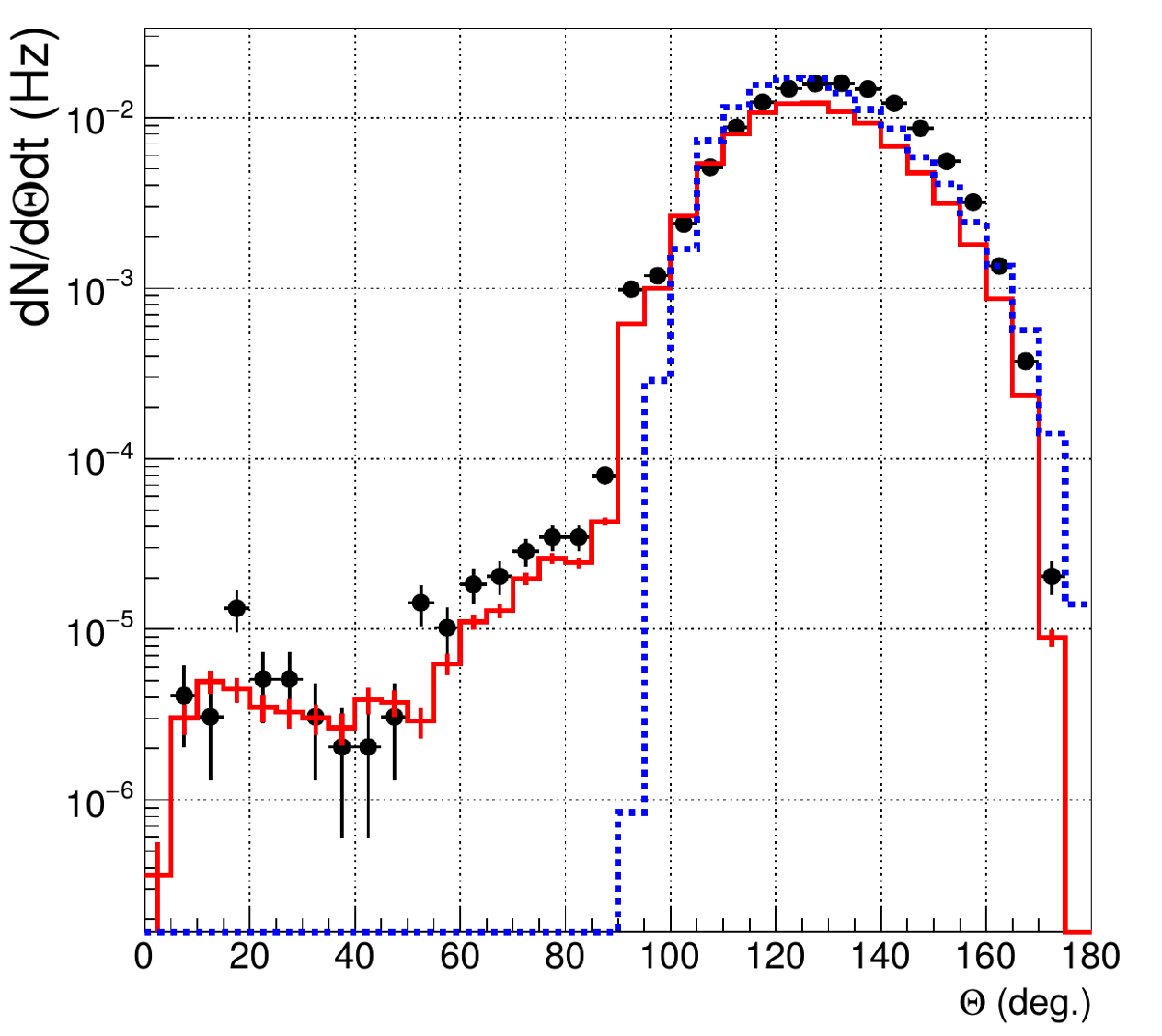} &
\includegraphics[width=75mm,height=60mm]{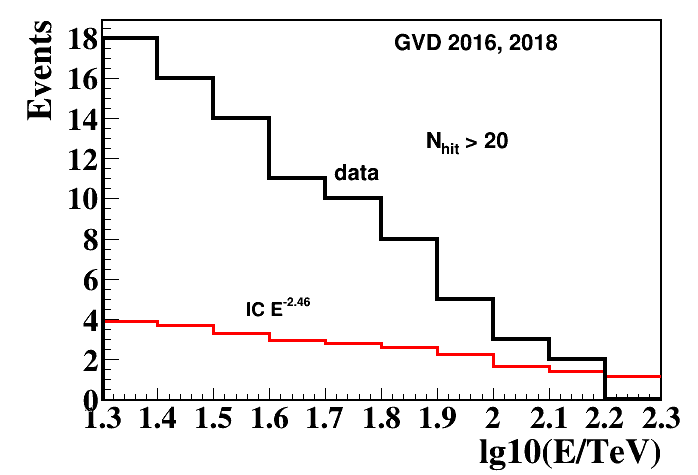}
\end{array}$
\end{center}
\caption{ Left panel: Zenith angle distribution reconstructed with data sample of atmospheric muon-like
  events  (black dots) and determined with Monte Carlo simulation of muons (red line) and upward moving muon
  neutrinos (blue line).
  Right panel: Cumulative energy distribution of experimental events (black histogram) end events
  expected from astrophysical flux with $E^{-2.46}$ energy spectrum and IceCube normalization (red
  histogram).} 
\label{fig.2}
\end{figure}

In Fig. \ref{fig.2} (left panel) the reconstructed angular distribution of muon-like events
deduced from the data sample is compared with Monte Carlo (MC) simulation of the detector
response to atmospheric muons and upward propagating muon neutrinos. Good agreement in shape and
rate between both muon distributions is achieved.

A procedure based on a boosted decision tree
(BDT) as implemented in the TMVA framework \cite{ricap14} was developed for the selection of neutrino events.
A set of quality variables was reconstructed for   each event and used for the BDT discriminant.
The BDT was trained on events reconstructed as upward going in MC
samples of atmospheric upward going neutrinos (signal) and atmospheric muons (background). Then
BDT value was reconstructed for the data events. Those events with BDT value larger than found
value for good discrimination of signal and background were selected. Overall, 23 neutrino candidate
events for 33 live days were found in data sample, while 42 events are expected from up-going
neutrino MC. Number of expected background from atmospheric muons is about 6. Improvements in
quality of track-like reconstruction is the next iteration of the data sample analysis.

\begin{figure}[!t]
\begin{center}$
\begin{array}{cc}
\includegraphics[width=75mm,height=80mm]{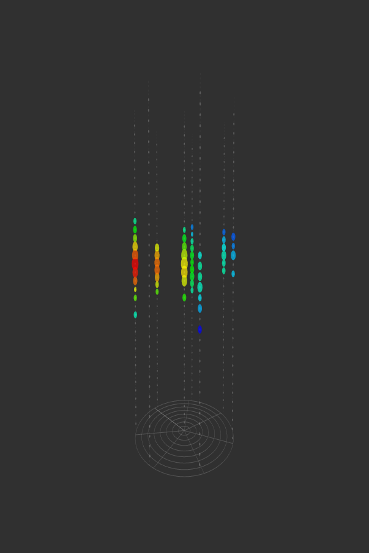} &
\includegraphics[width=75mm,height=80mm]{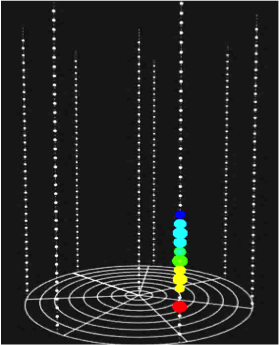} 
 \end{array}$
 \end{center}
\caption{Left panel: Cascade recorded on April 29, 2016 (see text). 
  Right panel: Reconstructed vertical event with 10 hit OMs.}
\label{fig.3}
\end{figure}

\subsection{Cascade detection by Baikal-GVD detector}

Currently, we do not have a clear theoretical idea about the sources of cosmic
neutrinos and how intense they are. We can categorize them as point and  sporadic
sources which might be related to high-energy processes in astrophysical objects and
$\gamma$-ray bursts, respectively. If  many  extragalactic  sources  contribute  to
an intense  and  measurable neutrino  emission we end up with a diffuse flux of
cosmic neutrinos. It can be related also to the interaction between cosmic rays and
Galactic gas/radiation fields resulting in the flux with $E^{-2.19}$ spectrum for energy
above 100 TeV. All these sources of cosmic neutrinos can be investigated with neutrino telescopes
with a volume about one kilometer cube or slightly smaller. A necessary condition
for it is that background processes in the neutrino detector are well understood. 

Cosmic ray interactions in the atmosphere produce neutrinos
at high energy. These constitute a diffuse, continuous and everywhere present flux,
consisting mostly of $\nu_\mu$, $\overline{\nu}_\mu$  with a spectrum distributed
as $E^{-3.7}_\nu$ in the region 100 GeV - 100 TeV  with some uncertainty
especially at high energies. Together with muons coming
from  above,  these  neutrinos  constitute   the  irreducible  background
when  searching  for cosmic neutrinos.

The exciting discovery of a diffuse flux of TeV to PeV neutrinos of undoubted
astrophysical origin was achieved with the cubic kilometer IceCube neutrino detector in 2013.
The obtained results demonstrate the importance of the cascade mode of neutrino detection
with neutrino telescopes. The Baikal collaboration has long-time experience in the search
for diffuse neutrino flux with NT200 telescope by using of cascade mode \cite{ricap11,ricap12}.
Baikal-GVD telescope has the potential to record astrophysical neutrinos with flux values
measured by IceCube even at early phases of construction by selecting 
cascade events generated by neutrino interactions in the sensitive volume of the detector.

To search for high-energy neutrinos of astrophysical origin the data collected by one cluster
in 2016 and by three clusters in 2018 have been used. A data sample of $3.8\times10^9$ events
has been accumulated by the array trigger, which corresponds to 872 one cluster live days.
After applying procedures of cascade vertex and energy reconstruction for hits with charge
higher than 1.5 ph.el., 417 cascade-like events with OMs hit multiplicity $N_{hit}>13$ have
been selected. The requirement of high hit multiplicity allows substantial suppression of
background events from atmospheric muon bundles. 18 of selected events have $N_{hit}>20$ and
3 of them where reconstructed with energies above 100 TeV and satisfy the requirements for
astrophysical neutrino selection. The calculations of the probability to obtain such high
multiplicity events from atmospheric muons and neutrinos are in progress. One of these three
events which was recorded on 29.04. 2016 is shown in Fig. 3 (left panel). Each sphere represents
an OM. Colors represent the arrival times of the photons where red indicates early and blue
late times. The size of spheres is a measure for the recorded number of photo-electrons.
Cumulative energy distribution of 18 experimental events with $N_{hit}>20$ (black histogram)
and distribution of
events expected from neutrino flux of astrophysical origin with power low energy spectrum and
IceCube normalization $1.7\times10^{-10}E^{-2.46}$ TeV$^{-1}$cm$^{-2}$s$^{-1}$sr$^{-1}$
(red histogram) are shown in Fig. 2 (right panel). For energies above 100 TeV 1.44 events are
expected from IC flux. In energy range below 100 TeV the data are
dominated by background events from atmospheric muons. For energies above 100 TeV higher
statistics is required for observation of the astrophysical neutrino flux.

Separately, a data analysis approach was developed for nearly vertical upward going muons, that
takes into account particular features of single string event reconstruction. Within the whole data sample
of 182 live days in 2016 there have been selected 5674 neutrino candidates. In this sample, there
was one candidate event with 10 hits, 6 events with 8 hits, 15 events with 7 hits and 144 events with
6 hits. No candidate events with 9 hits were found. The one event with 10 hits is shown in Fig. 3 (right panel).
Expected near vertical neutrino fluence can exceed number of background atmospheric neutrino
events as long as there can be an additional flux arising from the Earth's core. For the hypothesis of
dark matter in form of weakly interacting massive particles, WIMPs have been accumulated inside
the Earth during a time comparable to the universe age and currently weakly annihilate into ordinary
matter, producing high energy pairs of neutrinos and antineutrinos through decays of particles.

\subsection{Search for high-energy neutrinos associated with GW170817}

The most interesting source of variable luminosity of multimessenger interest was the galaxy NGC 4993, where
binary neutron stars merged and produced a gravitational wave GW170817, that was registered by the Advanced LIGO and
Advanced Virgo observatories \cite{qua11}. A short GRB (GRB170817A), associated with GW170817, was detected by Fermi-GBM
and INTEGRAL. Optical observations allowed the precise localization of the merger in the
galaxy NGC 4993 at a distance of $\sim$40 Mpc. High-energy neutrino signals associated with
the merger were searched for by the ANTARES and IceCube neutrino telescopes in muon
and cascade modes and the Pierre Auger Observatory \cite{qua12} and Super-Kamiokande \cite{qua13}. Two
different time windows were used for the searches. First, a $\pm$500 s time window around the
merger was used to search for neutrinos associated with prompt and extended gamma-ray
emission \cite{qua14,qua15}. Second, a 14-day time window following the GW detection, to cover
predictions of longer-lived emission processes \cite{qua16,qua17}. No significant neutrino signal was
observed by the neutrino telescopes.

\begin{figure}[!t]
\begin{center}$
\begin{array}{cc}
\includegraphics[width=75mm,height=60mm]{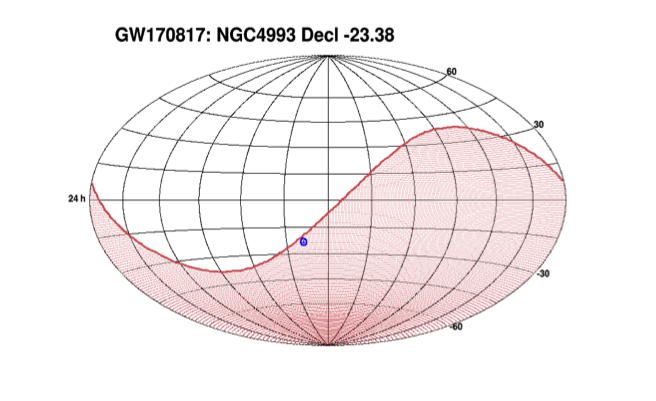} &
\includegraphics[width=75mm,height=60mm]{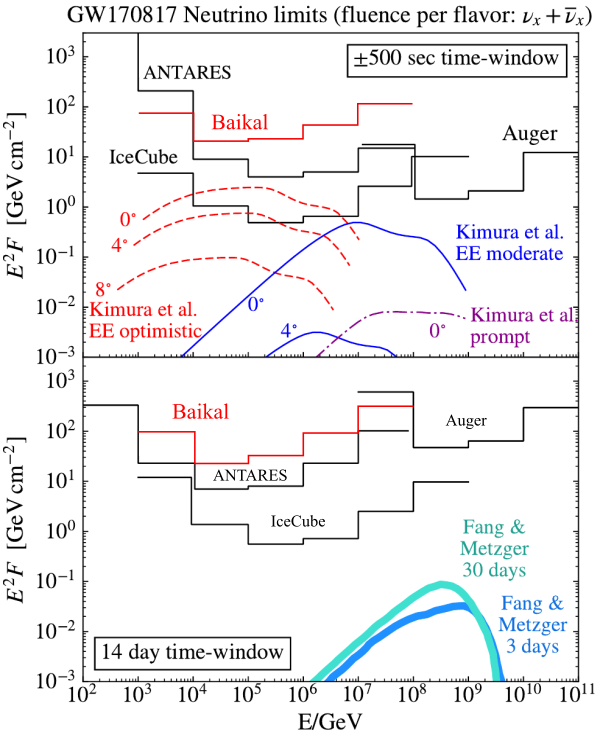} 
\end{array}$
\end{center}
\caption{Left panel: Horizon of the Baikal-GVD in alert time of the GW170817A (blue point).
  Right panel: The Baikal-GVD upper limits on neutrino fluences from direction of the GW170817A (see text).}
\label{fig.4}
\end{figure}

Two Baikal GVD-clusters were in operation in 2017.
Off-line analysis has been done in cascade mode to search for neutrino signal associated with
GW170817A.
In looking for fluence in prompt emission ($\pm$500 seconds) from GW170817A the
source was located slightly below the horizon for Baikal-GVD (zenith angle 93$^{\circ}$, horizon is shown
in Fig. \ref{fig.4} (left panel)). No
neutrino events associated with this source have been found in cascade search mode both in the
prompt emissions and in the emission delayed on 14 days after alert times. More details of analysis
are presented in Ref. \cite{ricap15}. Assuming $E^{-2}$ spectral behavior and equal fluence in neutrino flavors,
upper limits at 90\% c.l. have been derived on the neutrino fluence from GW170817A for each energy
decade and are shown in Fig. \ref{fig.4} (right panel).

\section{Conclusion}

An important progress has been achieved in the construction of the Baikal-GVD telescope by putting
in operation 5 clusters with effective volume of one fourth of the km$^3$. To reach an optimal performance
of the Baikal-GVD detector  advanced calibrations and analysis techniques have to be further
developed, tested and implemented in data processing. Currently, the identification of high energy neutrinos
is performed via muon tracks and cascade events in the detector and can be extended also to 
detection of double pulses later. By exploiting collected data sample of 2015-2018 years
the first high-energy cascade events were determined, which might be due to astrophysical
high energy neutrinos. In addition, the first neutrino-like events were reconstructed in muon track mode.
The attention was concentrated also on the search for neutrinos associated with GW170817A. It was concluded
that no high energy neutrino events from desired direction were registered by two Baikal-GVD clusters in
operation and an upper limits on the neutrino fluences were derived. With increasing
data records of steadily extending Baikal-GVD detector the importance of the largest fresh water
neutrino telescope for multimessenger study of astrophysical objects is growing. There are no doubts that Baikal-GVD telescope
will offer a unique view on our universe and provide powerful insights
into the performance of some of the most energetic and enigmatic objects in the cosmos.

\section{Acknowledgements}

This work was supported by  the Russian Foundation for Basic Research (Grants 16-29-13032, 17-0201237).

\end{document}